\documentclass[12pt]{article}
\usepackage{graphicx,amsmath,amssymb,color,bm}
\renewcommand{\baselinestretch}{2}

\begin{document}
\title{Unusual Coulomb excitations in ABC-stacked trilayer graphene \\}
\author{Chiun-Yan Lin, Ming-Hsun Lee, Ming-Fa Lin$^{*}$ \\
\small Department of Physics, National Cheng Kung University,
 Tainan, Taiwan\\
}

\maketitle



\renewcommand{\baselinestretch}{2}
\maketitle
\begin{abstract}
The layer-based random-phase approximation is further developed to
investigate electronic excitations in tri-layer ABC-stacked graphene. All the layer-dependent atomic interactions and Coulomb interactions are included in the dynamic charge screening. There exist rich and unique (momentum, frequency)-excitation phase diagrams, in which the complex single-particle excitations and five kinds of plasmon modes, are dominated by the unusual energy bands and doping carrier densities. The latter frequently experience the significant Landau damping due to the former, leading to the coexistence/destruction in the energy loss spectra. Specifically, the dispersion of the only acoustic plasmon in pristine case is dramatically changed from linear into quadratic even at very low doping.

\end{abstract}
\par\noindent ~~~~$^*$Corresponding author- E-mail: mflin@mail.ncku.edu.tw (M. F. Lin)

\vskip0.6 truecm

$\mathit{PACS}$: 81.05.U-, 78.67.Pt, 71.70.Di

\vskip 0.6 truecm

Few-layer graphenes are one of the main-stream 2D materials since the first discovery by the mechanical exfoliation on Bernal graphite in 2004.$^{1}$ Such systems have the unique geometric structures,  nano-scaled thickness, honeycomb lattice with two sublattices,  layered structure and  distinct stacking configurations. They are very suitable for studying the diverse physical phenomena, such as, the massless/massive Fermions,$^{2}$ the Coulomb excitations/deexcitations,$^{3-11}$ the quantized Landau levels,$^{11-14}$ the magneto-optical selection rules,$^{15-17}$ and the quantum Hall effects.$^{18-20}$ Specifically, the electronic excitations arising from the electron-electron (e-e) interactions  play critical roles in the energy and width of quasiparticle states. This work is focused on the rich Coulomb excitation spectra of the ABC-stacked trilayer graphene. The relationship between the momentum-frequency
phase diagram and the Fermi energy is investigated in detail.

Up to now, the stacking configurations, which are identified in the synthesized graphene systems, cover ABC,$^{21-24}$ ABA,$^{23}$ AAB,$^{25}$ and AAA.$^{26-27}$ They are the critical factor in determining the low-energy essential properties, e.g., the 2p$_z$-orbital-induced $\pi$-electronic structures. Among them, the ABC stacking, being predicted to have the lowest ground state energy,$^{28}$ is frequently observed in the experimental syntheses. This system presents very rich band structures under the various vertical and non-vertical interlayer atomic interactions. For example, the tri-layer ABC stacking possesses three pairs of weakly dispersive, sombrero-shaped and linear energy bands (Fig. 1(a)). Such energy dispersions will be directly reflected in other physical properties, e.g., the optical spectra$^{15,24}$ and low-frequency plasmon modes.


There are a lot of theoretical$^{3-8,11}$ and experimental$^{9,10}$ studies on the Coulomb excitations of graphene-related systems. The single-particle and collective excitations (SPEs and plasmons) are very sensitive to the stacking configurations, the number of layers, the electric and magnetic fields, and the dimensions. An intrinsic monolayer graphene only possesses the interband single-particle excitations (SPEs) at zero temperature because of the zero-gap semiconductor.$^{29}$ Three plasmon modes are revealed in the bilayer AA stacking, but not the bilayer AB stacking.$^{5,6}$ That the former has the sufficiently high free carrier density due to the interlayer atomic interactions accounts for this important difference. As for extrinsic few-layer graphenes, the doping free carriers can create the rich SPEs and plasmon modes.$^{4-6}$ However, most of the theoretical predictions only consider the electronic excitations arising from the first pair of valence and conduction bands nearest to the Fermi level ($E_F$). The fully dynamic charge screening due to all the pairs of energy bands will be included in the current calculations, so that the diverse Coulomb excitation spectra could be presented in momentum- and frequency-dependent phase diagrams.

For the ABC-stacked trilayer graphene, band structure and Coulomb interactions are, respectively, evaluated from the tight-binding model and random-phase approximation (RPA). Specially, the intralayer $\&$ the interlayer atomic interactions and Coulomb interactions are fully taken into consideration; furthermore, the layer-based polarization functions and dielectric functions are built from the sublattice-dependent tight-binding functions. How many kinds of SPE channels and plasmon modes will be explored in detail, especially for the strong dependence of electronic excitations on the magnitude of transferred momentum ($q$) and $E_F$. The predicted results could be verified from the high-resolution electron-energy-loss spectroscopy (EELS)$^{30-33}$ and inelastic light scattering spectroscopy.$^{34}$

The ABC-stacked trilayer graphene, as shown in Fig. 1(a),  has the significant interlayer atomic interactions (${\beta_1}$-${\beta_5}$) in addition to the intralayer one ($\beta_0$).$^{35}$ The former creates the layer-dependent Coulomb excitation behaviors. The $\pi$-electronic Hamiltonian is built from six 2$p_z$-dependent tight-binding functions. There are three pairs of valence and conduction bands, corresponding to the weakly dispersive ($S^{c,v}_{1}$), sombrero-shaped ($S^{c,v}_{2}$) and linear ($S^{c,v}_{3}$) dispersions, as shown in Fig. 1(b). The first pair belongs to surface-localized states, since they mainly come from the top and bottom layers. The electronic structures of ABC-stacked trilayer graphenes have been verified by ARPES$^{36}$, especially the presence of the surface-localized states clarified for the partially flat subbands centered at the K point. This will induce the unusual electronic excitations, compared with other stacking systems.  Each wave function is composed of six sublattice-based tight-binding functions, indicating the theoretical framework of the layer-dependent RPA.

When an electron beam is incident on the ABC-stacked trilayer graphene, the charge density distribution is assumed to be uniform inside each layer. The $\pi$ electrons on distinct layers will screen the time-dependent external potentials (${V_{ll^\prime}({\bf q})}$'s; the $l$-th layer) ) by the e-e interactions, leading to the induced charges and potentials. Within the linear response, the induced charge density is proportional to the effective Coulomb potentials (${V^{eff}_{ll^\prime}({\bf q},\omega\,)}$'s; $\omega$ the transferred frequency during charge screening). By using the layer-based RPA, the relationship among the effective, external and induced Coulomb potentials is characterized by the Dyson equation

\begin{equation}
\epsilon_{0}V^{eff}_{ll^{\prime}}(\mathbf{q},\omega)=V_{ll^{\prime}}(\mathbf{q})+
\sum\limits_{m,m^{\prime}}V_{lm}(\mathbf{q})P^{(1)}_{m,m^{\prime}}(\mathbf{q},\omega)
V^{eff}_{m^{\prime}l^{\prime}}(\mathbf{q},\omega)\text{,}
\end{equation}%
where $\epsilon_{0}(=2.4)$ is the background dielectric constant.$^{6}$

Apparently, the induced potential in the third term reveals the complicated  dynamic screening due to the intralayer and interlayer Coulomb interactions. The layer-dependent bare polarization function, being determined by energy bands and wave functions, is expressed as

\begin{equation}
\begin{array}{l}
P^{(1)}_{mm^{\prime}}(\mathbf{q},\omega)=2\sum\limits_{k}\sum\limits_{n,n^{\prime}}\sum\limits_{h,h^{\prime}=c,v}
\biggl(\sum\limits_{i}u^{h}_{nmi}(\mathbf{k})u^{h^{\prime}*}_{n^{\prime}m^{\prime}i}(\mathbf{k+q})\biggr)\\
\times\biggl(\sum\limits_{i^{\prime}}u^{h*}_{nm^{\prime}i^{\prime}}(\mathbf{k})
u^{h^{\prime}}_{n^{\prime}m^{\prime}i^{\prime}}
(\mathbf{k+q})\biggr)\times\frac{f(E^{h}_{n}(\mathbf{k}))-f(E^{h^{\prime}}_{n^{\prime}}(\mathbf{k+q}))}
{E^{h}_{n}(\mathbf{k})-E^{h^{\prime}}_{n^{\prime}}(\mathbf{k+q})
+\hbar\omega+i\Gamma}
\text{.}
\end{array}
\end{equation}%

${u^h_{nmi}}$ is the amplitude of wave function on the $i$th sublattice of the $m$-th layer, arising from the valence/conduction state  (${h=c \& v}$) of the $n$-th energy band. $f(E^{h}_{n}(\mathbf{k}))=1/[1+(E^{h}_{n}(\mathbf{k})-\mu(T))/k_{B}T]$ is the Fermi Dirac distribution. $k_{B}$, $\mu(T)$ and $\Gamma$ stand for the Boltzmann constant, chemical potential and the energy width due to various deexcitation mechanisms, respectively. Moreover, the layer-dependent dielectric function is defined from the linear relationship between the effective and external potentials:

\begin{equation}
\epsilon_{ll^{\prime}}(\mathbf{q},\omega)=\epsilon_{0}\delta_{ll^{\prime}}-
\sum\limits_{m}V_{lm}(\mathbf{q})P^{(1)}_{m,l^{\prime}}(\mathbf{q},\omega)\text{.}
\end{equation}%

Its inverse is useful in understanding the inelastic scattering probability  of the EELS measurement, the dimensional energy loss function:

\begin{equation}
\mathbf{Im}[-1/\epsilon]\equiv\sum\limits_{l}\mathbf{Im}\biggl[-V_{ll}(\mathbf{q},\omega) \biggr]
/\biggl(\sum\limits_{l,l^{\prime}}V_{ll^{\prime}}(q)/3\biggl)\text{.}
\end{equation}%

Equations (1)-(4), which cover all the atomic and Coulomb interactions, are suitable for any layered graphene systems.


The dynamic Coulomb response displays SPEs and collective excitations as the transferred $q$ and $\omega$ are conserved during e-e interactions. These two types of excitations are, respectively, characterized by the bare response function $P^{(1)}_{ll^{\prime}}(\mathbf{q},\omega)$ and energy loss function Im[-$\frac{1}{\epsilon}$]. The former, $P^{(1)}_{ll^{\prime}}(\mathbf{q},\omega)$, describes the dynamic charge screening and directly reflects the main features of the band structure. As shown in Figs. 1(c)-1(j), the interlayer polarizations $(l=l^{\prime})$ and intralayer polarizations $(l\neq l^{\prime})$ exhibit special structures associated with the critical points in the energy bands. In response to the dynamic Coulomb potential, the real part Re$[P^{(1)}_{ll^{\prime}}(\mathbf{q},\omega)]$ and imaginary part Im$[P^{(1)}_{ll^{\prime}}(\mathbf{q},\omega)]$ are linked to each other via the Kramers-Kronig relations.
The latter represents the strength of the SPEs, responsible for the Landau damping; its divergent singularities correspond to the Van Hove singularities in the joint density of states. In the ABC-stacked trilayer graphene, the $3\times3$ polarization function $P^{(1)}_{ll^{\prime}}(\mathbf{q},\omega)$ depends on the symmetry of the wavefunction on each layer. Revealing the corresponding excitations, the intralayer and interlayer polarizations have similar structures while their signs are determined by the phases of the wavefunctions. It is deduced that  $P^{(1)}_{11}=P^{(1)}_{33}$, $P^{(1)}_{12}=P^{(1)}_{23}$ and $P^{(1)}_{11}\simeq|P^{(1)}_{13}|$ due to the geometric inversion symmetry in the ABC configuration. For $E_{F}=0$ (black curves in Figs. 1(c)-1(f)), interband excitations give rise to divergent singularities of Im$P^{(1)}_{ll^{\prime}}(\mathbf{q},\omega)$ (indicated by the dashed gray lines). For those from $S^{v}_{1}\rightarrow S^{c}_{1}$ and $S^{v}_{2}\rightarrow S^{c}_{2}$, the square-root peaks from quasi-1D SPEs appear as a result of the nearly isotropic energy dispersions near the K point.$^{15}$ The others from $S^{v}_{1}\rightarrow S^{c}_{2}$ ($S^{v}_{2}\rightarrow S^{c}_{1}$) and $S^{v}_{2}\rightarrow S^{c}_{3}$ ($S^{v}_{3}\rightarrow S^{c}_{2}$) exhibit logarithmic form and display relatively weak response. In particular, the surface-localized states play an important role to the low-energy polarizations. Near the Fermi level, the prominent square-root divergent structures of Im$[P^{(1)}_{11}(\mathbf{q},\omega)]$ arise from the major low-energy excitations on the outmost layers, while the empty Im$[P^{(1)}_{22}(\mathbf{q},\omega)]$ demonstrates the absence of excitations on the middle layer. Based on the Kramers-Kronig relations, the square-root and logarithmic peaks in Re$[P^{(1)}_{ll^{\prime}}(\mathbf{q},\omega)]$ correspond to the square-root and step discontinuities in Im$[P^{(1)}_{ll^{\prime}}(\mathbf{q},\omega)]$.

When the Fermi level is higher, the free carriers play an important role in the low-energy excitations. More electronic excitation channels are triggered with the increasing free carriers under the influence of the interlayer atomic interactions and Coulomb interactions. Consequently, this leads to complicated polarization functions. At $E_{F}=0.1$ eV (figs. 1(c)-1(f)), the interlayer and intralayer polarizations have a similar structure, in which the first logarithmic singularity of Im$[P^{(1)}_{ll^{\prime}}(\mathbf{q},\omega)]$, shifting to higher $\omega$ with $q$, is mainly dominated by the SPEs within the $S^{c}_{1}\rightarrow S^{c}_{1}$ intraband region. Such channel determines the low-frequency excitation spectrum. On the other hand, the electronic states excited from $S^{c}_{1}$ subband induce new SPEs reaching up to $\simeq0.8$ eV (within the original interband region). It is claimed that when the energies of these SPEs coincide with those of plasmons, the plasmon intensity is weakened due to the Landau damping in the vicinity of the interband and intraband SPEs (dashed gray lines). When the Fermi level is increased from 0.3 eV to 0.8 eV (Figs. 1(g)-1(j)), the polarization functions obviously display strong responses and the intraband components gradually get more predominant than the interband ones. This implies that due to the interplay between interband and intraband excitations, the electronic excitation spectra can be diversified, and various plasmon modes are presented with a variation of $q$ and $E_{F}$.

The energy loss function Im[-$\frac{1}{\epsilon}$] is used to describe collective excitations, as shown in Fig. 2. For $E_{F}=0$ and $q=0.005{\AA}^{-1}$, there are two plasmon peaks, labeled by $\omega_{p}^{1st}$ and $\omega_{p}^{2nd}$, in the screened excitation spectrum of the pristine ABC-stacked trilayer graphene (black curve in Fig. 2(a)). Identified from the specified interband channel, i.e., $S^{v}_{1}\rightarrow S^{c}_{1}$, the plasmon energies correspond to the weak Landau damping, given by the imaginary parts of the bare response function in Fig. 1.
Responsible for the high density of states of the partially flat subbands, such interband plasmon mode with $\omega$ up to 0.25 eV is classified as the first kind of plasmons, $\omega^{1st}_{p}$. In the region near $\omega\simeq0.32$ eV, the intensity decrease of the loss spectrum is attributed to the Landau damping that matches the energies of the $S^{v}_{1}\rightarrow S^{c}_{2}$ SPEs. Modulated by the electron doping level, the plasmon intensity is enhanced for $E_{F}=0.1$ eV by the collective excitations within both intraband and interband channels (blue curve in Fig. 2(a)). There are three sharp plasmon modes, $\omega_{p}^{1st}$, $\omega_{p}^{2nd}$ and $\omega_{p}^{3rd}$, as identified by the excitation channels. The first plasmon mode $\omega_{p}^{1st}(\simeq0.1)$ eV is attributed to the $S^{c}_{1}\rightarrow S^{c}_{1}$ intraband excitation channel, leading to a relatively prominent plasmon intensity.
The latter two modes $\omega_{p}^{2nd}$ and $\omega_{p}^{3rd}$ near $0.3$ eV mainly come from the $S^{c}_{1}\rightarrow S^{c}_{2}$ and $S^{v}_{1}\rightarrow S^{c}_{1}$ interband excitations, respectively; however, the higher intraband ones also make considerable contributions to the collective excitations. Under the screen effects, the plasmons decline and broaden with an increment of $q$ and $\omega$. i.e., the plasmons decay into SPEs. When $E_{F}$ increases from 0.3 eV to 0.8 eV (Fig. 2(b)), the induced excitations lead to a dramatic change of the plasmon modes, as the doping level is higher than the critical points of the subbands $S^{c}_{2}$ and $S^{c}_{3}$ (green curves in Fig. 2(b)).
At $E_{F}=0.4$ eV, the first one peak is largely suppressed, implying the significant Landau damping resulting from SPEs. On the other hand, there are two new types of plasmon modes, $\omega_{p}^{4th}$ and $\omega_{p}^{5th}$, which are ascribed to the multi-mode excitations within various intraband and interband channels. Under a sufficiently large $E_{F}$, e.g., $E_{F}$=0.5, and 0.8 eV, there exists only one prominent peak, $\omega_{p}^{5th}$, of which the intensity and frequency are highly dependent on the densities of the free carriers. It should be noticed that for most interband excitations, plasmons and SPEs can coexist in a certain ($q,\omega$) region. The dispersion of each plasmon mode is confined by the boundaries of interband and intraband SPEs. The details are discussed in Figs. 3 and 4.


Trilaye ABC-stacked graphene exhibits rich and unique plasmon spectra in the ($q,\omega$)-excitation phase diagram. Various collective plasmon modes are presented in Figs. 3 and 4 under the influence of dynamic Coulomb interactions. In general, plasmons usually appear in specified domains of the $(q,\omega)$ diagram, because the Landau dampings occur in the region where the plasmon dispersion overlaps the continuum spectrum of electron-hole pairs (solid and dashed curves). The dispersion relation of the intrinsic plasmons $\omega_{p}^{1st}$ and $\omega_{p}^{2nd}$ are plotted as a function of $q$ in Fig. 3 (a), where the interband SPEs create strong Landau damping near $\omega\sim0.35$ eV and $\sim0.65$ eV. In particular, the first plasmon $\omega_{p}^{1st}$ is assigned to an acoustic mode, which approaches to a linear dependence on $q$ as a consequence of the collective excitation modes of the surface-localized states.
The linear plasmon dispersion, well defined up to 0.25 eV, is describable by the band-structure effect. Distinct from the $\sqrt{q}$ dispersion of the 2D electron gas and from that of the monolayer graphene, such a plasmon displays strong damping and disappears at small $q\simeq0.01{\AA}^{-1}$ (the SPE boundary of $S^{v}_{1} \rightarrow S^{c}_{2}$ and $S^{v}_{1} \rightarrow S^{c}_{3}$). After this region, the optical plasmon $\omega_{p}^{2nd}$ is formed near $\omega\simeq$ 0.32 eV, with the plasmon dispersion similar to the $\omega_{p}^{1st}$ one. These two plasmon modes have a similar dispersion which is mainly attributed to the $S_{1}^{v}-S_{1}^{c}$ interband collective excitation channel. Another prominent characteristic of the $\omega^{2nd}_{p}$ mode is the frequency reaching up to 0.6 eV. This is mainly based on the high DOS of $S_{1}^{v}$ and $S_{1}^{c}$ subbands that prevent the coupling from other interband excitations.




The plasmon modes are improved by doping to increase the free charge density in the extrinsic
condition.
As the Fermi level is increased, the interband and intraband excitations lead to new plasmon modes and diversified phase diagrams. At $E_{F}=0.1$ eV, the $\pi$ plasmons exhibit different dispersion relationships with the transferred momentum $q$. They behave as acoustic and optical modes in the low and middle ($q,\omega$) regions enclosed by the SPE boundaries (Figs. 3(b) and 3(c)). The acoustic mode is prominent in the region without SPEs, while shows strong damping when dispersing into the region of the $S^{v}_{1} \rightarrow S^{c}_{1}$ interband SPEs. Its intensity quickly drops by more than one order of magnitude at $q\simeq0.017{\AA}^{-1}$ and disappears beyond $q\simeq0.05{\AA}^{-1}$, a characteristic being dominated by the vertical nearest interlayer atomic interaction $\gamma_{1}$. On the other hand, the optical mode is separated into several parts, each of which appears with different degrees of Landau damping in a specified domain. For 0.3 eV $\leq\omega\leq$ 0.4 eV, the dispersion of the optical plasmon is approximately flat, reflecting the particular partially flat subbands. In addition to the original interband channels, the induced free carriers also contribute to the optical plasmon.

In the low-energy, the acoustic plasmon is deserved a closer examination. With an increment of the Fermi level, the collective excitation channels are transformed from interband ($S^{v}_{1} \rightarrow S^{c}_{1}$) to intraband ($S^{c}_{1} \rightarrow S^{c}_{1}$). Accordingly, the acoustic plasmon deviates from the liner dispersion of the pristine graphene even in the case of weak doping, as shown in Figs. 3(d)-3(g). The dispersion and intensity of the acoustic plasmon are enhanced, because the intraband collective excitations gradually become predominant in the plasmon spectra. Furthermore, the acoustic plasmon extends over a wider ($q,\omega$) range than in the case of zero doping as the SPE boundaries shift to higher $q$ and $\omega$.
The existence of the acoustic plasmons with different dispersion relationships indicates the effects on the band structure and the doping carrier densities.



With a variation of $E_{F}$, phase diagrams are dramatically changed due to the conservation of transferred momentum $q$ and energy $\omega$, as shown in Fig. 4.
At $E_{F}$=0.3 eV (Fig. 4(a)), the plasmon modes extend to higher energy due to the increasing free carriers. The most striking behavior of the $\omega_{p}^{1st}$ acoustic mode is its enhanced intensity and quadratic dispersion, which are in sharp contrast to the acoustic plasmon in cases of zero and low dopings. Nevertheless, if the subbands $S^{c}_{2}$, and $S^{c}_{3}$ are partially occupied, the plasmon modes are drastically changed. At $E_{F}$=0.4 eV (Figs. 4(b)), the quadratic acoustic plasmon arises from the three kinds of intraband excitations. i.e., $S^{c}_{i} \rightarrow S^{c}_{i}$ (i=1,2 and 3).
In addition, the interplay between interband and intraband excitations gives rise to new plasmon modes and diversified phase diagrams. According to the band effects, the Landau damping is strong for the induced interband SPEs, e.g., $S^{c}_{1} \rightarrow S^{c}_{2}$ and $S^{c}_{1} \rightarrow S^{c}_{3}$. In the region of 0.2 eV $\leq\omega\leq$ 0.3 eV, the $\omega_{p}^{4th}$ plasmon is classified as an abnormal plasmon mode with a concave upward dispersion; the weak plasmon intensity indicates the robust SPEs associated with the particular band structures in the ABC-stacked graphene, i.e., the partially flat and Sombrero subbands. The abnormal plasmon has an onset energy of negative dispersion about 0.3 eV and disperses upward for $q\gtrsim 0.01{\AA}^{-1}$. On the other hand, the $\omega_{p}^{5th}$ mode is enhanced and shifted to higher $\omega$ by the induced collective excitation channels.
With a further increase of $E_{F}$, the plasmon is hardly affected by the Landau dampings associated with the induced interband SPEs.
At $E_{F}$=0.5 eV (Fig. 4(c)), the various plasmons gradually merge into a long range acoustic mode, $\omega_{p}^{5th}$, because the collective excitation channels from the free carriers dominate the electronic excitations.
Under a heavy doping condition, e.g., $E_{F}$=0.8 eV (Fig. 4(d)), there exists only one strong acoustic mode, $\omega_{p}^{5th}$, over a wide region in the ($q, \omega$)-excitation phase diagram.

Tri-layer ABC-staked  graphene is predicted to exhibit the rich and unique Coulomb excitations. There are a lot of SPE channels and five kinds of plasmon modes, mainly arising from three pairs of energy bands and doping carrier densities. Their complicated relations create the diverse (momentum, frequency)-excitation phase diagrams. The plasmon peaks in the energy loss spectra might decline and even disappear under various Landau dampings. The linear acoustic plasmon is related to the surface states in pristine system, while it becomes an quadratic acoustic mode at any doping. Specially, all the layer-dependent atomic interactions and Coulomb interactions have been included in polarization function and dielectric function. The theoretical framework of the layer-based RPA could be further generalized to study the e-e interactions in emergent 2D materials, e.g., few-layer silicene and germanene.$^{37,38}$

\centerline {\textbf {ACKNOWLEDGMENTS}}%

\noindent
This work was supported in part by the National Science Council of Taiwan,
the Republic of China, under Grant Nos. NSC 105-2112-M-006 -002 -MY3.

\newpage
{\Large\bf References}
\renewcommand{\baselinestretch}{1}
\begin{itemize}

\item[${[1]}$]
K. S. Novoselov, A. K. Geim, S. V. Morozov, D. Jiang, Y. Zhang, S. V. Dubonos, I. V. Grigorieva, and A. A. Firsov, Science. $\mathbf{306}$, 666 (2004).

\item[${[2]}$]
A. H. Castro Neto, F. Guinea, N. M. R. Peres, K. S. Novoselov, and A. K. Geim, Rev. Mod. Phys. $\mathbf{81}$, 109 (2009).

\item[${[3]}$]
J. Y. Wu, S. C. Chen, O. Roslyak, G. Gumbs, and M. F. Lin, ACS Nano $\mathbf{5}$, 1026 (2011).

\item[${[4]}$]
T. Stauber, and H. Kohler, Nano Lett., $\mathbf{16}$, 6844 (2016).

\item[${[5]}$]
R. Roldan, and L. Brey, Phy. Rev. B $\mathbf{88}$, 115420 (2013).

\item[${[6]}$]
J. H. Ho, C. L. Lu, C. C. Hwang, C. P. Chang, and M. F. Lin, Phys. Rev. B $\mathbf{74}$, 085406 (2006).

\item[${[7]}$]
T. Ando, J. Phys. Soc. Jpn. $\mathbf{75}$, 074716 (2006).

\item[${[8]}$]
B. Wunsch, T. Stauber, F. Sols, and F. Guinea, New J. Phys. $\mathbf{8}$, 318 (2006).

\item[${[9]}$]
S. J. Park, and R. E. Palmer, Phys. Rev. Lett. $\mathbf{105}$, 016801 (2010).

\item[${[10]}$]
F. J. Nelson, J.-C. Idrobo, J. D. Fite, Z. L. Mi$\check{s}$kovi$\acute{c}$, S. J. Pennycook, S. T. Pantelides, J. U. Lee, and A. C. Diebold, Nano Lett., $\mathbf{14}$, 3827 (2014).

\item[${[11]}$]
M. O. Goerbig, Rev. Mod. Phys. $\mathbf{83}$, 1193 (2011).

\item[${[12]}$]
C. Y. Lin, J. Y. Wu, Y. J. Ou, Y. H. Chiu, and M. F. Lin, Phys. Chem. Chem. Phys. $\mathbf{17}$, 26008 (2015).

\item[${[13]}$]
C. Y. Lin, J. Y. Wu, Y. H. Chiu, C. P. Chang, and M. F. Lin, Phy. Rev. B $\mathbf{90}$, 205434 (2014).

\item[${[14]}$]
M. Koshino, and E. McCann, Phys. Rev. B $\mathbf{80}$, 165409 (2009).

\item[${[15]}$]
C. Y. Lin, T. N. Do, Y. K. Huang, and M. F. Lin, Optical properties of graphene in magnetic and electric fields, IOP Concise Physics. San Raefel, CA, USA: Morgan $\&$ Claypool
Publishers, 2017.

\item[${[16]}$]
Y. H. Ho, Y. H. Chiu, D. H. Lin, C. P. Chang, and M. F. Lin, ACS Nano $\mathbf{4}$, 1465 (2010).

\item[${[17]}$]
Z. Jiang, E. A. Henriksen, L. C. Tung, Y. J. Wang, M. E. Schwartz, M. Y. Han, P. Kim, and H. L. Stormer, Phys. Rev. Lett. $\mathbf{98}$, 197403 (2007).


\item[${[18]}$]
A. Kumar, W. Escoffier, J. M. Poumirol, C. Faugeras, D. P. Arovas, M. M. Fogler, F. Guinea, S. Roche, M. Goiran, and B. Raquet, Phys. Rev. Lett. $\mathbf{107}$, 126806 (2011).

\item[${[19]}$]
K. S. Novoselov, A. K. Geim, S. V. Morozov, D. Jiang, M. I. Katsnelson, I. V. Grigorieva, S. V. Dubonos, and A. A. Firsov, Nature $\mathbf{438}$, 197 (2005).

\item[${[20]}$]
L. Zhang, Y. Zhang, J. Camacho, M. Khodas, and I. Zaliznyak, Nat. Phys. $\mathbf{7}$, 953 (2011).

\item[${[21]}$]
R. Xu, L. J. Yin, J. B. Qiao, K. K. Bai, J. C. Nie, and L. He, Phys. Rev. B $\mathbf{91}$, 035410 (2015).

\item[${[22]}$]
D. Pierucci, H. Sediri, M. Hajlaoui, J. C. Girard, T. Brumme, M. Calandra, E. Velez-Fort, G. Patriarche, M. G. Silly, G. Ferro, V. Souliere, M. Marangolo, F. Sirotti, F. Mauri, and A. Ouerghi, ACS Nano $\mathbf{9}$, 5432 (2015).

\item[${[23]}$]
Y. Que, W. Xiao, H. Chen, D. Wang, S. Du, and H. -J. Gao, Appl. Phys. Lett. $\mathbf{107}$, 263101 (2015).

\item[${[24]}$]
K. F. Mak, J. Shan, and T. F. Heinz, Phys. Rev. Lett $\mathbf{104}$, 176404 (2010).

\item[${[25]}$]
L. B. Biedermann, M. L. Bolen, M. A. Capano, D. Zemlyanov, and R. G. Reifenberger, Phys. Rev. B $\mathbf{79}$, 125411 (2009).

\item[${[26]}$]
S. Horiuchi, T. Gotou, M. Fujiwara, R. Sotoaka, M. Hirata, K. K. Kimoto, T. Asaka, T. Yokosawa, Y. Matsui, K. Watanabe, Jpn. J. Appl. Phys. $\mathbf{47}$, 1073 (2003).

\item[${[27]}$]
J. Borysiuk, J. So${\l}$tys, and J. Piechota, J. Appl. Phys. $\mathbf{109}$, 093523 (2011).

\item[${[28]}$]
N. T. T. Tran, S. Y. Lin, C. Y. Lin, and M. F. Lin, Geometric and electronic properties of graphene-related systems: Chemical bondings, CRC Press, 2017.

\item[${[29]}$]
P. R. Wallace, Phys. Rev. $\mathbf{71}$, 622 (1947).

\item[${[30]}$]
M. D. Kapetanakis, W. Zhou, M. P. Oxley, J. Lee, M. P. Prange, S. J. Pennycook, J. C. Idrobo, and S. T. Pantelides, Phys. Rev. B $\mathbf{92}$, 125147 (2015).

\item[${[31]}$]
S. C. Liou, C.-S. Shie, C. H. Chen, R. Breitwieser, W. W. Pai, G. Y. Guo, and M.-W. Chu,
Phys. Rev. B $\mathbf{91}$, 045418 (2015).

\item[${[32]}$]
J. Lu, K. P. Loh, H. Huang, W. Chen, and A. T. S. Wee, Phys. Rev. B $\mathbf{80}$, 113410 (2009).

\item[${[33]}$]
P. Wachsmuth, R. Hambach, M. K. Kinyanjui, M. Guzzo, G. Benner, and U. Kaiser, Phys. Rev. B $\mathbf{88}$, 075433 (2013).

\item[${[34]}$]
C. F. Chen, C. H. Park, B. W. Boudouris, J. Horng, B. Geng, C. Girit, A. Zettl, M. F. Crommie, R. A. Segalman, S. G. Louie, and F. Wang, Nature $\mathbf{471}$, 617 (2011).

\item[${[35]}$]
J.-C. Charlier, J.-P. Michenaud, and Ph. Lambin, Phys. Rev. B $\mathbf{46}$, 4540 (1992).

\item[${[36]}$]
C. Coletti, S. Forti, A. Principi, K. V. Emtsev, A. A. Zakharov, K. M. Daniels, B. K. Daas, M. V. S. Chandrashekhar, T. Ouisse, D. Chaussende, A. H. MacDonald, M. Polini, and U. Starke,
Phys. Rev. B  $\mathbf{88}$, 155439 (2013).

\item[${[37]}$]
Jhao-Ying Wu, Szu-Chao Chen, Godfrey Gumbs, and Ming-Fa Lin,
Phys. Rev. B $\mathbf{94}$, 205427 (2016).

\item[${[38]}$]
P. H. Shih, Y. H. Chiu, J. Y. Wu, F. L. Shyu, and M. F. Lin, Sci. Rep. $\mathbf{7}$, 40600 (2017).



\end{itemize}

\newpage
{\Large\bf Figure captions}
\renewcommand{\baselinestretch}{1}
\begin{itemize}

\item[Figure 1:] (a) Geometric structure and (b) low-energy bands of ABC-stacked trilayer graphene. The Fermi level of the pristine graphene is set to be zero. (c) Imaginary and (d) real parts of $P^{(1)}_{11}$ at $E_{F}=0$ and $E_{F}=0.1$ eV for different $q's$. (e) and (f) correspond to those of $P^{(1)}_{22}$. (g)-(j) are related plots for $q=0.01$ at various $E_{F}'s$.

\item[Figure 2:] Energy loss spectra for (a) different $q's$, and (b) different $E_{F}'s$.

\item[Figure 3:] (a)-(c) ($q,\omega$)-excitation phase diagram of ABC-stacked trilayer graphene for $E_{F}=0$ and 0.1 eV. (d)-(g) Low-energy plasmons for $E_{F}$=0.01, 0.03, 0.05 and 0.07 eV. The boundaries of SPE channels are shown by solid and dashed curves, indicating the onset and end energies of the intraband and interband transitions.

\item[Figure 4:]($q,\omega$)-excitation phase diagram of ABC-stacked trilayer graphene at (a) $E_{F}=0.3$ eV, (b) $E_{F}=0.4$ eV, (c) $E_{F}=0.5$ eV and (d) $E_{F}=0.8$ eV.

\end{itemize}

\end{document}